\documentclass[12pt]{iopart}
\usepackage{graphicx}
\usepackage{iopams} 
\begin{document}
\title[Some remarks on the coherent-state variational
]{Some remarks on the coherent-state variational approach to
nonlinear boson models
}

\bigskip
\author{
P Buonsante 
and V Penna
}

\address{
Dipartimento di Fisica and Unit\`a C.N.I.S.M., Politecnico di Torino,
C.so Duca degli Abruzzi 24, I-10129 Torino, Italia}

\begin{abstract}
The mean-field pictures based on
the standard time-dependent variational approach have been
widely used in the study of nonlinear many-boson systems
such as the Bose-Hubbard model.
The mean-field schemes relevant to Gutzwiller-like trial
states $|F \rangle$, number-preserving states $|\xi \rangle$ and 
Glauber-like trial states $|Z \rangle$ are compared to evidence 
the specific properties of such schemes. After deriving the Hamiltonian
picture relevant to $|Z\rangle$ from that based on $|F\rangle$, the latter
is shown to exhibit a Poisson algebra equipped with a Weyl-Heisenberg
subalgebra which preludes to the $|Z\rangle$-based picture. Then states
$|Z \rangle$ are shown to be a superposition
of $\cal N$-boson states $|\xi \rangle$ and the
similarities/differences of the $|Z \rangle$-based and
$|\xi \rangle$-based pictures are discussed. Finally,
after proving that the simple, symmetric state $|\xi \rangle$
indeed corresponds to a SU(M) coherent state, a dual version
of states $|Z \rangle$ and $|\xi \rangle$ in terms of
momentum-mode operators is discussed together with
some applications. 
\end{abstract}




\section{Introduction}
\label{sez0}

The semiclassical formulation of many-mode boson models based on
the coherent-state (CS) method \cite{Per}, \cite{ZFG} has proven
to be an effective tool to describe the behavior of interacting 
bosons in many situations \cite{SolPe}-\cite{JAIN}.
Such models, usually represented by a second-quantized Hamiltonian in
terms of boson operators $a_i$, $a^+_i$, $n_i = a^+_i a_i$ 
with standard commutators $[a_m, a^+_i]= \delta_{mi}$,
exhibit a dramatic complexity owing to their many-body nonlinear character.
The combination of the CS method with the application of standard variational
schemes allows one to circumvent this problem by
reformulating model Hamiltonians into
a mean-field (${\cal MF}$) picture \cite{AP}, \cite{brown} 
%
in which the Schr\"odinger problem for variational
trial states $| \Phi \rangle = | \phi_1, \phi_2\, ...  \rangle$ describing the system quantum state 
is reduced to a set of Hamilton equations governing the evolution 
of parameters $\phi_j$. 
A very standard choice \cite{GIL} is 
$|\Phi\rangle \equiv |Z \rangle = \prod_i |z_i \rangle$
where, for each mode $a_i$, state $|z_i \rangle$ is a Glauber
CS satisfying the defining equation $a_m |z_m \rangle
= z_m |z_m\rangle$ and $\phi_j$ identify with CS parameters
$z_j= \langle \Phi | a_j |\Phi \rangle$. 
Similar schemes have been developed for magnetic and multi-level atomic 
systems \cite{AP}, \cite{Arec} where $|\Phi \rangle$ is a product of spin CS.
The important feature is that dynamical variables $\phi_j$ are, 
at the same time,
the expectation values of the Hamiltonian operators and thus
are providing information on microscopic physical processes.
Significant examples are found within the physics of ultracold bosons 
confined in optical lattices where this $\cal MF$ formulation
successfully describes complex dynamical behaviors \cite{BOB}-\cite{BPV}. 
There, CS parameters usually represent either the 
on-site superfluid order parameters of local condensates
or the expectation values of local operators such as $a^+_i a_m$ and
$n_i$ giving informations on space correlations and
boson populations, respectively.

The most general version of this $\cal MF$ picture, however, 
is achieved by using a Gutzwiller-like \cite{GUTZ} trial state
$|\Phi\rangle$ exhibiting yet a factorized semiclassical form,
but involving constituent states more structured than CS. These are
\begin{equation}
\label{GTS}
|F\rangle = \prod^M_{i=1} |F_i \rangle = 
\prod^M_{i=1} \sum^{\infty}_{n_i=0} f^i_{n_i} | n \rangle_i
\, ,
\end{equation}
where $|F_i \rangle$ has replaced local state $|z_i \rangle$ in $|Z\rangle$, 
$n_i | n \rangle_i = n | n \rangle_i$ and $M$ is the boson-mode number namely,
for many models, the lattice-site number. This choice ensures an improved
description of microscopic processes in the sense that, for each mode,
infinitely-many variational parameters $f^i_{n_i}$ are available now in place 
of the $M$ parameters $z_i$ of $|Z \rangle$. The $|F \rangle$-based approach
has been applied in studying the dynamics of Bose-Hubbard (BH) model \cite{JVC},
\cite{DST} as well as its zero-temperature critical properties
\cite{SAC}-\cite{lasp}.

Recently, a third variational scheme has been considered
\cite{LAK}, \cite{BPV} to approach the dynamics of many-mode boson models, 
where state $|\Phi \rangle$ is assumed to have the form 
\begin{equation}
|{\cal N}, \xi \rangle =
({{\cal N}!})^{-1/2}\, (A^+)^{\cal N} |0, 0, ...,0 \rangle
\, , 
\label{suMCS}
\end{equation}
where $A^+ = \sum^M_{i=1} \xi_i a^+_i$ and the constraint $\sum^M_{i=1} |\xi_i|^2=1$ ensuring its normalization.
Different from states $|Z \rangle$ and $|F \rangle$, the distinctive property
\cite{com1} of states $|\xi \rangle$  (${\cal N}$ will be often implied in
$|{\cal N}, \xi \rangle$) is to diagonalize, by construction, the boson-number
operator $N= \sum_i n_i$ whose eigenvalue can be easily shown to identify with
index ${\cal N}$. Hence, states $| \xi \rangle$ having ${\cal N}$ as a good
quantum number naturally embody the property $[N, H]= 0$ characterizing usually
many-mode boson Hamiltonians $H$. This valuable feature reflects in turn an even
important fact that states \ref{suMCS} actually coincide with the CS of group
SU(M) where eigenvalue $\cal N$ is the index labeling the representation of SU(M). 

The structure of formula \ref{suMCS}, however, appears quite different from
the (standard) group-theoretic form of SU(M) coherent states. The standard
definition \cite{Per}, in fact, states that $|\xi \rangle = g | \Omega \rangle$
with $g \in$ SU(M) where $|\Omega\rangle$ is an appropriate extremal state.
Then a CS is generated through the exponential action of an algebra element
$a \in$ su(M) such that $g =\exp (i a)$ where $a$ is in general a linear
combination of su(M) generators. A well-known $M$ boson-mode realization
\cite{ZFG} of SU(M) CS is, for example,
$$
g |\Omega\rangle = T(\zeta) |\Omega\rangle
\, , \quad
|\Omega\rangle= 
%
|0, \, ... ,{\cal N},0, ... \rangle
$$
where $T(\zeta)=e^{ia}$ is the displacement operator, 
$a= \! \sum^M_{\ell \ne m}
( \zeta^*_\ell a^+_m a_\ell +\zeta_\ell a^+_\ell a_m )$,
$\zeta_\ell \in {\bf C}$,
and $n_\ell |\Omega\rangle = \delta_{\ell m} {\cal N}|\Omega\rangle$. Such a
definition has been (and is) currently in use to represent quantum dynamical
processes in microscopic systems within Quantum Optics, Condensed-Matter theory,
and Nuclear Physics (see \cite{ZFG}, \cite{KLASKA} and references therein).

Except for the case $M=2$, where states \ref{suMCS} are easily related \cite{ZFG}
to definition $g | \Omega \rangle$ (see \ref{M2}), for $M \ge 3$ the connection of
formula \ref{suMCS} with the group-theoretic form of SU(M) CS is less direct owing
to the difficulty in disentangling the group element $g$ through
the Baker-Campbell-Hausdorff decomposition \cite{ZFG}.
As noted in \cite{N1}, where this issue has been investigated, even if SU(2) CS
have been widely used in the literature, not much work has been devoted to find
realizations of SU(M) CS more practicable in physical applications.
In this respect, indeed definition \ref{suMCS} has proven to supply a useful
tool in applications. However, despite states \ref{suMCS} being tacitly presented
as CS of group SU(M) in various papers, their connection with the group-theoretic definition within CS Theory is far from being evident. We thus devote some attention
to this particular aspect even if it might be  known to authors 
\cite{N1}-\cite{SUN3}, \cite{Vou} involved on mathematical aspects of CS.

This paper is aimed at comparing the $\cal MF$ schemes based on states $|F \rangle$,
$|Z \rangle$ and $|\xi \rangle$, widely used in applications to boson systems. We
enlight some formal aspects concerning both the implementation of the
(time-dependent) variational approach within such schemes, and the representation
of trial states in terms of CS. We emphasize that some parts of our discussion have
a review character and involve well-known theoretic tools.
Nevertheless, a direct comparison among these three $\cal MF$ schemes
has never been presented to our knowledge. We feel that such a comparison
can elucidate their specific properties and advantages at the operational
level prompting as well their applications.

A first objective of this paper is to evidence how the variational schemes based
on state $|F \rangle$ and $|Z \rangle$, respectively, are related to each other.
In section \ref{sez2}, after reviewing the variational procedure that amounts to reformulating quantum Hamiltonian models in terms of effective Hamiltonians 
and of the corresponding dynamical equations, we focus on the formal derivation
of the Hamiltonian picture relevant to $|Z\rangle$ from the one based on
$|F\rangle$. We show that the Poisson algebra of variables ${f}^{j}_n$,
${\bar f}^{\ell}_m$ is naturally equipped with a (classical) Weyl-Heisenberg
sub-algebra preluding to the  $|Z\rangle$-based picture.

A second objective is to relate SU(M) CS $|\xi \rangle$
to states $|Z \rangle$ and $|F \rangle$. In section \ref{sez3},
following \cite{GIL}, we show that Glauber-like trial state
$$
|Z \rangle = \prod^M_{i=1} |z_i \rangle 
\, , \qquad
|z_i \rangle = e^{z_i a^+_i- z^*_i a_i} |0 \rangle_i
= e^{-|z_i|^2/2} \sum^{\infty}_{n=0} \frac{z^n_i}{\sqrt {n!} } 
|n \rangle_i 
\, , \quad
z_i \in {\bf C}
$$
can be expressed 
as a superposition of SU(M) states $|\xi \rangle$ 
thus making evident why the
dynamical equations obtained in the $|Z \rangle$-based scheme
have essentially the same form of those obtained in the 
$|\xi \rangle$-based scheme. To illustrate this situation 
we derive the $\cal MF$ dynamics relevant to the BH model
showing (see \ref{tre}) how the use of form \ref{suMCS} in place 
of the standard SU(M) CS definition is extremely advantageous.
In this section we also display 
an explicit way to relate state \ref{suMCS} with the standard form
$|\xi \rangle = g | \Omega \rangle$ of CS theory
involving one among $M$ possible (equivalent) choices of 
extremal vector $| \Omega \rangle$ and of the relevant maximal isotropy algebra.
Finally, in section \ref{sez4}, we discuss the property of ``duality"
inherent in space-like states $|Z \rangle$ and $|\xi \rangle$
defined in terms of ambient-lattice operators $a^+_i$ showing how they can be
easily rewritten as momentum-like states involving momentum modes $b^+_k$.
We exploit this property to construct Schr\"odinger cat states in terms of
states $|\xi \rangle$ and to show that they exhibit specific momentum features.

\section{Mean-field approaches based on states $|F \rangle$ and $|Z \rangle$}
\label{sez2}

In order to compare the $|Z \rangle$-based approach
with the $|F \rangle$-based approach we refer to the $\cal MF$
dynamical equations stemmed from such schemes for
the well-known Bose-Hubbard Hamiltonian \cite{AP}-\cite{KYP}.
Optical-lattice confinement shows that real boson lattice 
systems are effectively described within the Bose-Hubbard
picture. The corresponding Hamiltonian \cite{Fisher}, defined
on an $M$-site lattice, is
\begin{equation}
H = \frac{U}{2} \sum_j  (n_j^2 -n_j) - 
%
%
\sum_{<j \ell >} \, T_{j \ell } \,  a_j^+ a_\ell 
\label{BHH}
\end{equation}
where $n_i= a^+_i a_i$ ($i=1,..., M$) 
and $a_i$, $a^+_i$ obey the standard commutators 
$[a_m, a^+_i]= \delta_{mi}$.
In the hopping term $\sum_{<j \ell >} \equiv 
\frac{1}{2} \sum_j \sum_{\ell \in j}$, where $\ell$ labels the nearest neighbour sites of $j$, and $T_{j \ell } = T_{\ell j}$.
This model well represents the boson tunneling among the potential wells forming $d$-dimensional arrays ($d = 1,2,3$) through the hopping amplitude 
$T_{j \ell }$, and takes into account boson-boson interactions
by means of parameter $U$. For one-dimensional homogeneous arrays 
the hopping term 
reduces to $T \Sigma_j ( a_{j+1}^+ a_j + a_{j+1} a^+_j)$.

The $\cal MF$ dynamics relevant to
trial state  \ref{GTS} is easily derived. 
State $|F \rangle = \Pi_i |F_i \rangle$,
where $|F_i \rangle = \sum_n f^i_n |n \rangle_i $ and $|n \rangle_i$
is such that $a^+_i |n \rangle_i= {\sqrt{1+n}} |n+1 \rangle_i$,
$a_i |n \rangle_i= {\sqrt{n}} |n-1 \rangle_i$,
obeys the normalization condition
$\langle F | F \rangle
= \Pi_i \left [\,  \sum_n | f^{i}_n |^2 \right ] = \, 1$.
The application of the time-dependent variational principle 
\cite{ZFG}, \cite{AP} amounts to
deriving dynamical equation for parameters $f^i_{n}$ by stationarizing
the weak form of Schr\"odinger equation 
$\langle \Psi | S | \Psi \rangle = 0$ 
where $S := i\hbar \partial_t -H$.
In order to illustrate this procedure, we write explicitly the expectation
value of microscopic physical operators appearing in $H$. These are
\begin{equation}
\alpha_i = \langle F | a_i |F \rangle =
{\sum}_m {\sqrt {m }} \, \langle F_i | f^{i}_{m} |m-1 \rangle 
=
{\sum}_m {\sqrt {m +1}} \,  {\bar f}^{i}_m f^{i}_{m+1} \, ,
\label{ev1}
\end{equation}
%
%
%
$\langle F | a_i a^+_j |F \rangle
= \langle F_i | a_i |F_i \rangle \,
\langle F_j |  a^+_j |F_j \rangle
= \alpha_i \alpha_j^*$, 
and
$\langle F | (n_i)^s |F \rangle= \sum_n \, n^s  | f^{i}_n |^2 $
where the exponent $s$ is an integer.
To calculate $\langle \Psi | \partial_t | \Psi \rangle$
and ${\cal H}=\langle \Psi | H | \Psi \rangle$ 
in $\langle \Psi | S | \Psi \rangle$
we standardly set
$|\Psi \rangle = e^{iA} |F \rangle$, the phase $A$ representing the
effective action within the variational procedure. The first quantity becomes
$\langle \Psi | \partial_t | \Psi \rangle=
i{\dot A} + \langle F | \partial_t |F \rangle$ with
$$
\langle F | \partial_t |F \rangle
%
%
= {\sum}_j \, {\sum}_n {\bar f}^{j}_n {\dot f}^{j}_n 
=
\frac{1}{2} {\sum}_j {\sum}_n \left [ {\bar f}^{j}_n \frac{ d f^{j}_n}{dt}
-\frac{d {\bar f}^{j}_n }{dt} f^{j}_n \right ]
+\frac{d}{dt} \, {\sum}_j {\sum}_n  | f^{j}_n |^2
$$
while ${\cal H}= \langle F | H |F \rangle$ reads
\begin{equation}
{\cal H} = 
\frac{U}{2} {\sum}_j \left ( {\sum}_n (n^2-n) |{f}^{j}_n|^2 \right )
- {\sum}_{<\ell j>} \, T_{\ell j} \, \alpha_\ell \alpha_j^*
\label{Ham1}
\end{equation}
From the action $A = \int L dt 
= \int  dt \, [ \, i \hbar \langle F | \partial_t |F \rangle 
- {\cal H} \, ]$ (in which the second term of
$\langle F | \partial_t |F \rangle$, a total time-derivative, 
can be eliminated) one obtains the Lagrange equations
$$
\frac{\partial}{\partial t} \frac{d L}{d {\dot {\bar { f^i_m } } } }
-\, \frac{d L}{d {{\bar f}^{i}_m } } = 0
\, ,
\quad
\frac{\partial}{\partial t} \frac{d L}{d {\dot {f}^{i}_m } }
-\, \frac{d L}{d {{f}^{i}_m } } = 0
\, ,
$$
($d f^i_m /dt = {\dot { f^i_m } }$) that can equivalently be written as
(we set $\hbar=1$)
\begin{equation}
-i {\dot f }^{i}_m  + 
\frac{\partial {\cal H} }{\partial {{\bar f}^{i}_m } } = 0
\, , \quad
i {\dot {\bar {f^{i}_m}} }  + 
\frac{\partial {\cal H} }{\partial {{f}^{i}_m } } = 0\, .
\label{lagham}
\end{equation}
By defining the Poisson brackets
\begin{equation}
\{ A, B  \} = - i\, {\sum}_n {\sum}_j \left [
\frac{\partial A }{\partial {{ f}^{j}_n } }
\frac{\partial B }{\partial {{\bar f}^{j}_n } }
-
\frac{\partial B }{\partial {{ f}^{j}_n } }
\frac{\partial A }{\partial {{\bar f}^{j}_n } }
\right ]
\label{PB1}
\end{equation}
equations (\ref{lagham}) can be formulated within the 
standard Hamiltonian formalism as
$ d f^i_m /dt = \{ f^{i}_m , {\cal H} \}$ and
$ d {\bar {f^{i}_m} }/dt = \{ {\bar f}^{i}_m , {\cal H} \}$.
%
%
The resulting $\cal MF$ dynamical equations are
\begin{equation}
i {\dot f}^{i}_m = \frac{U}{2} (m^2-m) f^{i}_m
-{\sqrt {m+1}} f^{i}_{m+1} \, \Phi^*_i 
-{\sqrt {m}} f^{i}_{m-1} \Phi_i \, , 
\label{Geq}
\end{equation}
(equations for ${\bar f}^{i}_m $ are obtained from the latter by
complex conjugation), where
$$
\Phi^*_i = 
{\sum}_{j \in i} \, T_{i j} \, \alpha_j^*
\, , \qquad
\Phi_i = {\sum}_{\ell \in i} \, T_{\ell i} \, \alpha_\ell
\, ,
$$
and, in addition to definition \ref{ev1}, one has
$
\alpha^*_i =\sum^{\infty}_{m=0} {\sqrt {m}}\, {\bar f}^{i}_{m} f^{i}_{m-1} 
$.
Concluding, we recall that within the $|F\rangle$-based scheme
the average total particle number 
${\cal N}= \langle F| N |F \rangle= \sum_i \langle F| n_i |F \rangle
= \sum_i\sum_n n |f^n_i|^2$ should be a constant of motion of Hamiltonian 
\ref{Ham1}. The validity of this property is verified in \ref{due}.
In addition to ${\cal N}$, other $M$ motion constants can
be shown to be
represented by $I_i =\sum_n |f^n_i|^2 = \langle F_i |F_i \rangle$.
These allows one to implement the local-state normalization
condition $\langle F_i |F_i \rangle=1$.

\subsection{\bf Connection with the Glauber-like trial state scenario}

A simple assumption allows one to recover state $|Z\rangle$ from 
$|F\rangle$ and relate the corresponding variational schemes. This is
\begin{equation}
f^i_m = e^{-|z_i|^2/2} z^m_i/\sqrt {m!}
\label{assum}
\end{equation}
entailing 
$\sum^{\infty}_{n=0} f^i_{n} | n \rangle_i=
e^{-|z_i|^2/2}  \sum^{\infty}_{n=0} {(a^+_i)^n} | 0 \rangle_i/ {n!}$ 
and therefore
\begin{equation}
|F \rangle = 
\prod^M_{i=1} \sum^{\infty}_{n=0} f^i_{n} | n \rangle_i
=
\prod^M_{i=1} e^{z_i a^+_i- z^*_i a_i} | 0 \rangle_i
=
\prod^M_{i=1} | z_i \rangle = | Z \rangle\, ,
\label{SFZ}
\end{equation}
where the defining formula of Glauber CS $| z \rangle = 
e^{z a^+- z^* a} | 0 \rangle = 
e^{-|z|^2/2}e^{z a^+} |0\rangle$ has been used for each space mode 
together with the Baker-Campbell-Hausdorf decomposition formula \cite{Per}.
The same assumption enables one to find the new form of
parameter \ref{ev1}
\begin{equation}
\alpha_j = 
\sum^{\infty}_{m=0} {\sqrt {m +1}} \, {\bar f}^{j}_m f^{j}_{m+1}
=
e^{-|z_j|^2} \sum^{\infty}_{m=0} \, z_j \,
\frac{|z_j|^{2m} }{m!} = \, z_j \, ,
\label{zeta}
\end{equation}
showing that $\alpha_j$ reduce to
Glauber CS parameters $z_j$, and formulas
$\langle F | a_i a^+_j |F \rangle= z_i z_j^*$,
$\langle F | n_i |F \rangle= | z_i |^2 $ and
$\langle F | n^2_i |F \rangle= | z_i |^4+| z_i |^2$.
In order to recover the $\cal MF$
equations inherent in the $|Z \rangle$-based picture
we consider the time derivative of $\alpha_j$. This is given by
$
i{\dot \alpha}_j = \sum^{\infty}_{m=0} {\sqrt {m +1}} \, 
[ i f^{j}_{m+1}\, {d {\bar f}^{j}_m }/{dt}
+ i {\bar f}^{j}_m\, {d f^{i}_{m+1} }/{dt} ]
$,
which reduces to (the detailed calculation is carried out in \ref{glauber})
\begin{equation}
i \frac{d \alpha_j}{dt} 
=
\sum^{\infty}_{m=0}  \, 
\Bigl [  U m {\sqrt {m +1}} \, {\bar f}^{j}_m f^{j}_{m+1} \Bigr ]
- \Phi_j \, .
\label{mot1}
\end{equation}
Notice that, as illustrated in \ref{glauber}, no explicit assumption on 
the form of $f^{j}_{m}$ has been requested so far (except for 
$\langle F_j|F_j \rangle =1$) in getting \ref{mot1}.
At this point, however, the use of formula \ref{assum} 
in \ref{mot1} becomes necessary. We find
$
\sum^{\infty}_{m=0}  \, U m {\sqrt {m +1}} \, {\bar f}^{j}_m f^{j}_{m+1}
=
$
$
U z_j \sum^{\infty}_{m=0}  \, m \, {\bar f}^{j}_m f^{j}_{m}= U z_j |z_j|^2
$
which leads, in turn, to the well-known final equations
\begin{equation}
\qquad \qquad
i {\dot {z_j} } = U  z_j |z_j|^2 - {\sum}_{\ell \in j} T_{\ell j} z_\ell\, ,
\label{eqZ}
\end{equation}
describing a set of discrete nonlinear Schr\"ondinger equations \cite{J},
namely the $\cal MF$ dynamical equations associated to the Bose-Hubbard
model within the Glauber-like variational picture \cite{AP}. 
Equations \ref{eqZ} can be issued from the new Hamiltonian
\begin{equation}
{\cal H} = \frac{U}{2} {\sum}_i |z_i|^4 - 
{\sum}_{<j\ell>} T_{j \ell} \, z_j z^*_\ell
\label{cBHH}
\end{equation}
obtained by rewriting formula \ref{Ham1} in terms of $z_j$, 
and defining the new Poisson brackets (PB)
\begin{equation}
\{ A, B  \} = -i\, {\sum}_\ell 
\left [
\frac{\partial A }{\partial {z_\ell }}
\frac{\partial B }{\partial {{\bar z}_\ell } }
-
\frac{\partial B }{\partial {z_\ell } }
\frac{\partial A }{\partial {{\bar z}_\ell } }
\right ] 
\quad
\Leftrightarrow 
\quad
\{ z_j, z^*_\ell  \} = -i \delta_{j \ell}
\, .
\label{PB2}
\end{equation}
The crucial point that explains and justifies
the whole reduction of the Hamiltonian picture 
based variables ${f}^{i}_n$, ${\bar f}^{\ell}_m$ to 
the one involving the restricted set of variables $z_i$, ${\bar z}_\ell$
thus consists in showing that PB $\ref{PB2}$ are consistent 
with PB of $\ref{PB1}$. In particular, $\alpha_i$, $\alpha^*_\ell$
must be shown to exhibit, within algebra $\ref{PB1}$,
the same algebraic structure of variables $z_j$, $z^*_\ell$.
By setting $A= \alpha_i$ and $B=\alpha^*_\ell$ in PB $\ref{PB1}$
one discovers
that
$$
\{\alpha_j, \alpha^*_\ell \}=
{\sum}_m {\sum}_n \sqrt{m+1} \sqrt{n+1} 
\left \{ {\bar f}^{j}_m {f}^{j}_{m+1},  
f^{\ell}_n {\bar f}^{\ell}_{n+1} \right \}
=
$$
\begin{equation}
=
-i \delta _{j \ell}
{\sum}_m (m+1)  \left ( |{f}^{j}_{m}|^2- |f^{j}_{m+1}|^2 \right )
= - i \delta _{j \ell} {\sum}_m |{f}^{j}_{m}|^2 = - i \delta _{j \ell}
\label{alfabra}
\end{equation}
due to normalization $\langle F_i |F_i \rangle=1$.
Also, one easily proves that $\{\alpha_j, N_\ell \}=
- i \delta _{j \ell} \alpha_j$, 
where $N_\ell = \langle F| n_\ell |F \rangle= \sum_n n |{f}^{\ell}_n|^2$.
Hence, it is an intrinsic feature of algebra \ref{PB1} characterized by
$\{ {f}^{j}_n,  {\bar f}^{\ell}_m \}= - i \, \delta_{j\ell} \delta_{nm}$
the property that $\alpha_i$, $\alpha^*_\ell$ form a (classical) Weyl-Heisenberg 
sub-algebra of algebra \ref{PB1}. Noticeably, the latter
represents the classical counterpart of 
the original boson algebra
$[a_j, n_\ell ]= \delta_{j \ell} a_j$, 
$[a_j, a^+_\ell ]= \delta_{j \ell}$ charaterizing Hamiltonian \ref{BHH}.
Then identities $\alpha_j\equiv z_j$, and $\alpha^*_\ell \equiv z^*_\ell $,
obtained by assuming ${f}^{j}_{m}$ as a function of $z_j$ (see formula \ref{zeta}), 
quite naturally entail that $z_j$, $z^*_\ell$ obey the canonical brackets given
in \ref{PB2} within the Glauber-like scheme. This completes the proof that the
$|Z\rangle$-based variational picture is consistently contained within
that based on the more structured state $|F\rangle$.

Concluding, we notice that if an effective Hamiltonian $\cal H$
depending on ${f}^{i}_n$, ${\bar f}^{\ell}_m$ can be rewritten
in terms of collective variables $\alpha_j$, $\alpha^*_\ell$
then sub-algebra \ref{alfabra} is sufficient for determining the
evolution of the system and the $|F\rangle$-based picture
becomes redundant. This is not the case of Hamiltonian \ref{Ham1} 
where, owing to the presence of the nonlinear $U$-dependent term, 
algebra \ref{PB1} is necessary to derive the relevant motion equations. 

Comparing equations \ref{Geq} and \ref{eqZ} fully evidences how the more
pronounced quantum character of the $|F\rangle$-based picture involves a
dynamical scenario of great complexity. The marked semiclassical character of
the $|Z\rangle$-based picture instead appears when comparing quantum model
\ref{BHH} with Hamiltonian \ref{cBHH}. 
The latter, in fact, is essentially obtained from \ref{BHH} through the 
substitutions $a_i \to z_i$, $a^+_i \to {\bar z}_i$, namely by implementing
the Bogoliubov approximation. At a formal level, the $|Z\rangle$-based scheme
thus provides an effective, dynamically-consistent formulation of the 
Bogoliubov semiclassical picture.

\section{Mean-field approach based on state $|\xi \rangle$}
\label{sez3}

A quite significant form of state $| Z \rangle$ given by 
\ref{SFZ} is achieved with a simple calculation
$$
| Z \rangle= 
\prod^M_{i=1} | z_i \rangle = 
e^{-\frac{1}{2} \sum_i |z_i|^2}
\prod^M_{i=1} e^{z_i a^+_i} | 0 \rangle_i
= e^{-\sum_i |z_i|^2/2} e^{\sum_i z_i a^+_i} | 0 \rangle_i
=
$$
$$
=
e^{-\frac{1}{2} \sum_i |z_i|^2} 
\sum^\infty_{S=0} \frac{1}{S!}
\left ( \sum_i z_i a^+_i \right)^S \, | 0, 0\, ...0 \rangle
=
e^{-{\cal N}/{2}} \sum^\infty_{S=0} 
\frac{ {{\cal N} }^{\frac{S}{2}} }{\sqrt {S!}}\, |S; \xi \rangle
\, ,
$$
where $\Pi^M_{i=1} | 0 \rangle_i= | 0, 0,\, ...0 \rangle$,
$\xi_i = z_i /{\sqrt {\cal N}}$,
and $|S; \xi \rangle$ corresponds to state $|\xi \rangle$ defined 
by \ref{suMCS} where the group-representation index $S$  
has been evidenced. State $|S; \xi \rangle$ is characterized by 
$\langle \xi; S|N |S; \xi \rangle = S$ and the orthogonality property
$\langle \zeta; S' |S; \xi \rangle = \delta_{SS'}$.
The final version of $|Z\rangle$ derives from the observation that
$\langle Z|N | Z \rangle= \sum_i |z_i|^2= {\cal N}$ is
the average total boson number in $| Z \rangle$-based scheme and that 
$\xi_i = z_i /{\sqrt {\cal N}}$ is consistent with
the normalization condition $\sum_i |\xi_i|^2 = 1$ of SU(M) coherent
states. The latter follows from the scalar-product formula of two CS states $|\xi \rangle$ and $|\eta \rangle$ given by
$\langle \eta |\xi\rangle =\left (\Sigma_{i} \eta^*_i \xi_i \right )^{S}$.

The new information about $|Z\rangle$ is therefore that states $|S; \xi \rangle$
are its constitutive elements. In particular, state \ref{suMCS} features the
property of incorporating only contributions of Fock states pertaining to the
$S$-particle sectors of the Hilbert space. This becomes quite evident rewriting
$|S;\xi \rangle$ as
\begin{equation}
|S; \xi \rangle = \frac{1}{\sqrt {S!}}
\left ( \sum_i \xi_i a^+_i \right)^{S} \, |0\rangle
=
\sum^{(S)}_{\vec m} \frac{\sqrt {S!}}{\sqrt{\prod_i (m_i!)} }
\, \xi^{m_1}_1\, ...\xi^{m_M}_M\, | \vec m \rangle
\label{Sxi}
\end{equation}
where $|0 \rangle= | 0, 0\, ...0 \rangle$ and superscript $(S)$ recalls that $S=\sum_i m_i$ 
and $| \vec m \rangle$ is such that
$$
| \vec m \rangle= | m_1, \, ... m_M \rangle
=\prod^{(S)}_i \frac{ (a^+_i)^{m_i}}{\sqrt{m_i!}}
 | 0,0\, ...0 \rangle
\quad
\Rightarrow
\quad
N |\vec m \rangle = \sum_i m_i |\vec m \rangle = S |\vec m \rangle\, .
$$
%
The previous formulas
allows one to evaluate the weight of state $|L ; \zeta \rangle $ 
in $|Z\rangle$
$$
\langle \zeta ; L | Z \rangle=
e^{-{\cal N}/{2}} \sum^\infty_{S=0} \frac{{\cal N}^{S/2}}{\sqrt {S!}}
\, \langle \zeta ; L |S; \xi \rangle
= e^{-\frac{{\cal N}}{2} } \frac{{\cal N}^{ \frac{L}{2}} }{\sqrt {L!}} 
\left (\sum_i \zeta^*_i \xi_i \right )^L
\, .
$$
Upon setting $\zeta =\xi$, the normalization condition $\sum_i |\xi_i|^2 =1$
entails that 
$\langle \zeta ; L | Z \rangle
= e^{-{\cal N}/2 }{{\cal N}^{L/2}}/{\sqrt {L!}}$ whose
maximum value is reached for $ L \equiv {\cal N}$ (${\cal N}$ is assumed 
to be integer). Considering $|L ; \xi \rangle $ with $L= {\cal N} \pm p$ 
and $p << L$, one easily finds that
the state-weight distribution around the maximum-weight state
$|{\cal N} ; \xi \rangle $ is not sharp.

The variational procedure reviewed in section \ref{sez2} can once more be
applied to BH model \ref{BHH} assuming $| \Psi \rangle = e^{iA} | \xi \rangle$
as the trial state. The weak form of Schr\"odinger equation 
$\langle \Psi | (i\hbar \partial_t -H) | \Psi \rangle = 0$ provides the action 
$A = \int dt {\cal L} (\xi)$ where the effective Lagrangian
${\cal L} (\xi) = 
i\hbar \langle \xi |\partial_t | \xi \rangle -\langle \xi | H| \xi \rangle$
supplies the dynamical equations of variables $\xi_i$. The calculation of both
$\langle \xi |\partial_t | \xi \rangle$ and
${\cal H} (\xi) = \langle \xi | H| \xi \rangle$
is carried out in \ref{tre} together with the basic formulas required
to achieve these results. We find, in particular, that the average local
boson number is $\langle \xi | n_i | \xi \rangle = {\cal N} |\xi_j|^2 $,
reproducing consistently  
$\langle \xi | N| \xi \rangle = {\cal N} \sum_i |\xi_j|^2 = {\cal N}$. 
This suggests to define variables $\psi_j = {\sqrt N} \xi_j$ 
(formally coinciding with $z_j$)
for better comparing the present $\cal MF$ dynamics with that
issued from trial state $|Z\rangle$. Explicitly, one finds
$\langle \xi | \partial_t |\xi \rangle = 
{\cal N}\sum_j {\dot \xi}_j \xi^*_j =\sum_j {\dot \psi}_j \psi^*_j$
and 
$$
{\cal H} (\xi) = \langle \xi | H| \xi \rangle = 
\frac{U ({\cal N}-1)}{2 {\cal N}} \sum_j  |\psi_j|^4 - 
\, \sum_{<j \ell>} T_{j \ell} \, \psi_j^* \psi_{\ell} 
$$
while the dynamics is found to be governed by 
\begin{equation}
i \frac{d \psi_j}{dt} = U \frac{({\cal N}-1)}{{\cal N}} |\psi_j|^2 \psi_j
- \sum_{j \in \ell} T_{j \ell} \psi_{\ell} \, .
\label{eqSUM}
\end{equation}
Equations \ref{eqSUM} can be interpreted as the projection of equations 
\ref{eqZ} on a given $S$-particle Hilbert-space sector. In order to prove this
property one has to consider the variational scheme based on a generic
state $|\psi \rangle = \sum_S C_S |S;\xi\rangle$. The latter reproduces
state $|Z \rangle$ when the condition
$C_S = (S!)^{-1/2} e^{-{\cal N}/2} {\cal N}^{S/2}$ is imposed.
The $|\psi \rangle$-based scheme would involve, in this case, the effective
Lagrangian 
${\cal L} = i\hbar \langle \psi |\partial_t | \psi \rangle -
\langle \psi | H| \psi \rangle$
which, in the case $|\psi \rangle=|Z \rangle$, leads to equations \ref{eqZ}.
Observing that $\langle \xi; R|S;\xi\rangle = \delta_{RS}$, then
$$
\langle \psi | X | \psi \rangle = {\sum}_R {\sum}_S
C_R C_S \langle \xi; R| X |S;\xi\rangle
= {\sum}_S C^2_S \langle \xi; S| X |S;\xi\rangle
\, ,
$$
for both $X = H$ and $X= \partial_t$,
states $\partial_t |S;\xi\rangle$ and $H|S;\xi\rangle$ pertaining to
the $S$-particle sector. Hence, ${\cal L}$ reduces to a summation 
${\cal L} = \sum_S C^2_S \, {\cal L}_S(\xi)$ over independent 
$S$-particle Lagrangians
$ {\cal L}_S(\xi)=\langle \xi; S|[ i\hbar\partial_t -H ]|S;\xi\rangle$,
the case $S = {\cal N}$ giving equations \ref{eqSUM}.

Formally, no significant difference therefore distinguishes equations
\ref{eqSUM} (and the relevant generating Hamiltonian) from the picture
corresponding to equations \ref{eqZ} if $({\cal N}-1)/{\cal N} \to 1$
namely for boson numbers ${\cal N}$ sufficiently large. A profound difference
instead concerns the role of variables $z_i$ and $\psi_i$ in the relevant schemes.
While $z_i= \langle Z | a_i|Z \rangle$ relates $a_i$ to the local
superfluid parameter $z_i$, its counterpart in the
$|\xi \rangle$-based scheme has no explicit physical meaning since
$\langle \xi | a_i|\xi \rangle=0$. State $a_i|\xi \rangle$ belongs
in fact to the $({\cal N}-1)$-particle Hilbert-space sector thus resulting othogonal to the ${\cal N}$-particle state $|\xi \rangle$. With state $|Z\rangle$ this effect is avoided since $|Z\rangle$ 
is spread on the whole Hilbert space.
The equivalence between the two schemes
is restored in the case of two-particle operators
$z_i z^*_j= \langle Z | a_i a^+_j|Z \rangle$ being comparable with
$\psi_{i} \psi_j^* = {\cal N} \xi_{i} \xi_j^* =
\langle \xi | a_i a^+_j|\xi \rangle$. Of course case $i=j$ 
describing local populations $\langle \Phi |n_i |\Phi\rangle$,  
$\Phi = Z, \xi$ is also included. Variables 
$\psi_j = |\psi_j| e^{i\theta_j}$ thus
acquire a physical meaning in terms of local populations
$\langle \xi |n_i |\xi\rangle= |\psi_{i}|^2$. The relevant
phases $\theta_j$ have no role unless one considers 
expectation values of operators $a_i a^+_j$ involving 
phase differences $\theta_i- \theta_j$.

\subsection{Group-theoretic form of state $|\xi \rangle$}
\label{sez3.1}

State \ref{suMCS} displays the particularly nice property 
to possess a fully symmetric structure mirroring the fact that
all modes $a_m$ play an equal important role
in defining $|\xi \rangle$. This symmetry must be ``broken'' for 
proving that \ref{suMCS} has the standard
group-theoretic CS form where $|\xi \rangle$ is generated by the
group action on an extremal state (the choice of the latter entails
the loss of the symmetric form).
To show this our first step consists in proving that formula \ref{suMCS} 
can be rewritten as
\begin{equation}
|\xi \rangle = ({\cal N}!)^{-1/2} 
(A^+)^{\cal N} |0, 0, ...,0 \rangle = 
({\cal N}!)^{-1/2} E\, (a_1^+)^{\cal N} E^+ |0, 0, ...,0 \rangle 
\label{equiv}
\end{equation}
where $E^+=E^{-1}$ and $E$ is an element of SU(M) whose parametrization in terms
of variables $\xi_i$ can be easily determined. The action of  $a^+_i$ on the
zero-boson Fock state $|0 \rangle := |0, 0, ...,0 \rangle$ is the standard one 
$(a^+_i)^p |0, 0, ...,0 \rangle = {\sqrt {n_i !}} |..., 0, n_i, ... \rangle $
with $n_i = p$ while $a_i \, |0 \rangle = 0$.
We point out that the choice of generating $A^+$ from $a_m^+$ rather than from
$a_1^+$ is equally possible and simply entails choosing, in turn, one from $M$
possible parametrizations for $|\xi \rangle$ and the relevant
form of $E$. 
This arbitrariness reflects the just mentioned symmetry of formula \ref{suMCS}.
For proving \ref{equiv} we show that
$E\, a_1^+ E^+ = A^+$ where $E = e^{iS}\, e^{iD}$ is defined as
$$
S= \sum^M_{i=1} \phi_i n_i \, , \quad
\quad D = \sum^M_{i=2} \, \theta_i (  a^+_1 a_i + a^+_i a_1) \, ,
\quad ({D^+} = D) 
$$
with $\phi_i \in {\bf R}$ and $\theta_i \in {\bf R}$.
Upon setting $\sum^M_{k=2} \theta^2_k := \theta^2$, standard calculations 
show that
$
e^{iD} a^+_1 e^{-iD} = \sum^M_{j=1} y_j  a^+_j
$
(see \ref{quattro}) 
with $y_1 = \cos \theta$ and $y_k= i\theta_k {\sin \theta}/{\theta}$
if $ k \ne 1$. A further action of $e^{iS}$ gives
\begin{equation}
e^{iS} \, e^{iD} \, a^+_1 e^{-iD} \, e^{-iS}=
\sum^M_{j=1} y_j e^{i \phi_j} a^+_j = \sum^M_{j=1} \xi_j a^+_j 
= A^+ \, ,
\label{A} 
\end{equation}
where $\xi_1 = e^{i \phi_1} \cos \theta$,
$\xi_k =  i \theta_k e^{i \phi_k} \sin \theta /\theta$ and the action of
factor $e^{i \phi_i n_i}$ in $e^{iS} = \Pi_i e^{i \phi_i n_i}$ is described by
$e^{i \phi_\ell n_\ell} a^+_\ell e^{-i \phi_\ell n_\ell}=
e^{i \phi_\ell} a^+_\ell $.
The identification $A^+ =E a_1^+ E^+$ suggested by formula \ref{A} 
is confirmed by the fact that
the correct normalization of $A^+$ components $\xi_j$
follows from
$
\sum^M_{j=1} |\xi_j|^2 = \cos^2 \theta + \sum^M_{k=2} 
({\theta^2_k}/{\theta^2}) \sin^2 \theta = 1
$.
Since $a^+_m a_i |0 \rangle = n_i |0 \rangle=  0$ 
and thus $S |0 \rangle = D |0 \rangle= 0$,
formula \ref{equiv} becomes
$$
|\xi \rangle= ({\cal N}!)^{-1/2}
E \, (a_1^+)^{\cal N} E^+ |0 \rangle 
=({\cal N}!)^{-1/2}
E \, (a_1^+)^{\cal N} |0 \rangle 
= E \, |{\cal N} , 0, ...,0 \rangle
\, ,
$$
being $E^+ |0 \rangle = e^{-iD} e^{-iS}|0 \rangle
= |0 \rangle$ and $(a_1^+)^{\cal N} |0 \rangle ={\sqrt {{\cal N}!}} 
|{\cal N}, 0, ...\rangle$. 
Upon observing that 
$E= e^{iS} e^{iD} = 
\exp \left [ e^{iS} \, {iD}\, e^{-iS} \right ] e^{iS}$ 
state $|\xi \rangle$ can be rewritten as
$$
|\xi \rangle =
 e^{i \phi_1 {\cal N}} \exp \left [ e^{iS} \, {iD}\, e^{-iS} \right ]\, 
|{\cal N}, 0, ...,0 \rangle
=
e^{i \phi_1 {\cal N}}\, T(\zeta) |{\cal N}, 0, ...,0 \rangle 
\, ,
$$
where 
$e^{iS} \, a^+_\ell a_1 \, e^{-iS} =e^{i(\phi_\ell-\phi_1)} a^+_\ell a_1$
entails that
$$
e^{iS}\, {D}\, e^{-iS} =\sum^M_{\ell =2}\, 
(\zeta^*_\ell a^+_1 a_\ell +\zeta_\ell a^+_\ell a_1)
\, ,
\quad
T(\zeta) := 
e^{i\sum^M_{\ell =2} \, ( \zeta^*_\ell a^+_1 a_\ell +\zeta_\ell a^+_\ell a_1)}
\, ,
$$
with $\zeta_\ell = \theta_\ell e^{i(\phi_\ell -\phi_1)}$,
$\ell = 2, 3\, ...,M$. Summarizing, we have found that
\begin{equation}
|\xi \rangle =
\frac{1}{\sqrt {{\cal N}!}} 
(A^+)^{\cal N} |0, 0, ...,0 \rangle
=
e^{i \phi_1 {\cal N}}\, T(\zeta) |{\cal N}, 0, ...,0 \rangle 
\, , 
\label{cinq}
\end{equation}
%
%
where $T(\zeta)$ is an element of group SU(M), which proves that state 
\ref{suMCS}, up to an irrelevant phase factor, is generated by 
the group action of $T(\zeta)$. The identification of $T(\zeta)$
with an element of SU(M) is discussed in \ref{uno}.
By setting $\phi_1 =0$,
the relation between $\xi_i$ and $\zeta_\ell$
is described by $\xi_\ell= \zeta_\ell \sin \theta /{\theta}$
whereas $\xi_1$ is fixed by $|\xi_1|^2= 1- \sum^M_{\ell =2} |\xi_\ell |^2$.
An initial choice with $A^+ = E a^+_m E^+ $ in formula \ref{equiv} 
would have entailed $T(\zeta)$ generated by 
$\sum^M_{\ell \ne m} \, 
( \zeta^*_\ell a^+_m a_\ell +\zeta_\ell a^+_\ell a_m )$
and the extremal state 
$|0,... {\cal N}, ... \rangle 
= ({\cal N}!)^{-1/2} (a^+_m)^{\cal N} |0 \rangle$.

As a final step, we prove that formula \ref{cinq} is consistent
with the group-theoretic definition of CS based on the
notion of maximal isotropy subalgebra (MIS). 
Within the CS Theory \cite{Per} a class of CS is 
derived by identifying the (complex) MIS $\cal B$
of $\cal G$ = su(M) and the related extremal vector $|\psi_0 \rangle$. 
The defining formula for $|\psi_0 \rangle$ states that 
$a |\psi_0 \rangle = \lambda_a |\psi_0 \rangle$, $\lambda_a \in {\bf C}$,
$\forall a \in {\cal G }_0 $ where
${\cal B } \cap {\cal G } := {\cal G }_0 $.
The (complex) MIS naturally related to
formula \ref{cinq} is given by
$$
{\cal B }= \{h_k ,\, a^+_1 a_k, \, a^+_\ell a_k \, (k \ne \ell)\, : 
k, \ell \in [2,M] \}
\, , \quad ([ {\cal B }, {\cal B }] \subseteq {\cal B })
$$ 
whose generators are such that 
$a^+_1 a_k |{\cal N}, 0, ...,0 \rangle = a^+_\ell a_k |{\cal N}, 0, ...,0 \rangle =0$
and generators $h_k$ form the Cartan (abelian) subalgebra.
The vector satisfying the defining formula for $|\psi_0 \rangle$ is thus
$|{\cal N}, 0, ...,0 \rangle$.
According to the MIS scheme, 
coherent states are generated by the action on $|\psi_0 \rangle$
of the elements of the quotient group
$G^c/ B$ where $G^c = \exp {\cal G}$
and $B = \exp {\cal B}$. The algebra that generates $G^c/ B$ is in our case
$\{ a^+_k a_1 \, : k \in [2,M] \}$,
which entails that a coherent state,  
up to a normalization factor $\lambda$, has the form
\begin{equation}
\lambda \, e^{\sum_k \eta_k a^+_k a_1} |{\cal N}, 0, ...,0 \rangle
\, , \quad e^{\sum_k \eta_k a^+_k a_1} \, \in G^c/ B\, .
\label{sei}
\end{equation}
State \ref{cinq} precisely has this form. In order to check this property, one
can observe that
$T(\zeta) = 
\exp \left [{i \sum^M_{k=2} \,  
( \zeta^*_k a^+_1 a_k + \zeta_k a^+_k a_1)} \right ]
= e^{i \theta ( a^+_1 D + D^+ a_1)}
$,
with $D = \sum^M_{k=2} \, {\zeta^*_k a_k }/{\theta} $
where
$$
[D, D^+ \, ] = \sum^M_{k=2} \sum^M_{\ell =2}
\,\frac{\zeta^*_k \zeta_\ell }{\theta^2}  \, [a_k , a^+_\ell ]
= \sum^M_{k=2} \, \frac{|\zeta_k|^2 }{\theta^2} = 1
\, .
$$
The exponent of $T(\zeta)$ can thus be viewed as 
an element of su(2) in the two-boson (Schwinger) realization 
with generators 
$J_- = a^+_1 D$, $J_+ = a_1 D^+$, $J_3 = (D^+ D- a^+_1 a_1 )/2$
and commutators 
$[J_3, J_{\pm}]= \pm J_{\pm}$ and $[J_{+}, J_{-}]= 2J_3$. 
This information allows us to apply the standard decomposition formula
$
e^{vJ_+ - v^*J_-} = e^{u J_+ }\, e^{{\rm ln} (1+|u|^2)}\, e^{- u^*J_-}
$
for the SU(2) elements \cite{Per} where 
$v, u \in {\bf C}$, $v= |v|e^{i \psi}$, $u= |u|e^{i \psi}$
and $|u| = {\rm tg} |v|$. Setting $v= i \theta$, which entails 
$u = i \, {\rm tg} \, \theta$,
one has $T(\zeta) = 
e^{i \theta (D^+_1 B + D^+ a_1)} = e^{i \theta  (J_- + J_+ )} $
thus obtaining
$$
T(\zeta) |{\cal N}, 0, ...,0 \rangle
=
\frac{e^{u J_+ }}{(1+|u|^2)^{\cal N}}  |{\cal N}, 0, ...,0 \rangle
= \frac{e^{ \sum^M_{k=2} \, \eta_k a^+_k a_1 
}}{(1+|u|^2)^{\cal N}}  |{\cal N}, 0, ...,0 \rangle
$$
where
$\eta_k = u \zeta_k /\theta = i e^{i\theta_k} {\rm tg} \theta$, 
and $J_3|{\cal N}, 0, ...,0 \rangle=
({\cal N}/2) |{\cal N}, 0, ...,0 \rangle $ has been used together with
$J_-|{\cal N}, 0, ...,0 \rangle= a_1^+ D |{\cal N}, 0, ...,0 \rangle= 0$.
Therefore state \ref{cinq} indeed can be cast into the CS form 
\ref{sei} determined within the theory of CS.
%
%
%
\section{The duality property of states $|Z \rangle$ and $|\xi \rangle $}
\label{sez4}

Both state $|Z \rangle$ and $|\xi \rangle $, whose definition involves
boson operators $a_j$ and $a^+_j$ of the ambient lattice, can be shown exhibiting
a dual character which becomes evident when space-like operators are expressed as momentum-like operators through Fourier formulas
\begin{equation}
b_q = \sum^M_{j=1} \frac{e^{-i{\tilde q} j}}{\sqrt M} a_j
\, ,\quad 
a_j = \sum^M_{j=1} \frac{e^{i{\tilde q} j}}{\sqrt M} b_q \, ,
\quad
{\tilde q} := 2\pi q/M \, , \, q\in [1,M]
\label{F1}
\end{equation}
where $[a_j, a^+_\ell ]= \delta_{j \ell} $ implies that 
$[b_q, b^+_p ]= \delta_{q p}$. Notice that we assume periodic boundary
conditions (namely the lattice is a closed ring) so that displacements 
$q \to q +sM$ and $j \to j+ rM$ ($r$, $s$ are integer) leave operators 
$a_j$ and $b_q$
unchanged. This condition, standardly assumed to simplify theoretical models, 
becomes necessary in real lattices with a ring geometry \cite{ring}. 
Concerning state $|Z \rangle = \prod_j |z_j\rangle$ simple calculations yield
\begin{equation}
|Z \rangle = e^{-\frac{1}{2} \sum_j |z_j|^2 } e^{\sum_j z_j a^+_j} |0 \rangle
= e^{-\frac{1}{2}\sum_k |v_k|^2} e^{\sum_k v_k b^+_k} |0 \rangle
= \prod_k |v_k\rangle = |V \rangle
\label{ZV}
\end{equation}
where $a_\ell |0 \rangle = 0 = b_k |0 \rangle$ has been used 
(recall that $|0 \rangle= |0, 0, ...,0 \rangle$) and,
thanks to definitions \ref{F1}, one has 
$v_k = \sum^M_{j=1} {e^{-i{\tilde q} j}} z_j/{\sqrt M}$,
and
$z_j = \sum^M_{j=1} {e^{i{\tilde q} j}} \, v_q/{\sqrt M}$.
Trial states $|Z \rangle$ are thus equivalent to states 
$|V \rangle$ formed by momentum-like Glauber CS $|v_k\rangle= 
e^{v_k b^+_k- v^*_k b_k} |0 \rangle_k $. Similarly, states
$| \xi \rangle$ transform into momentum-like SU(M) CS
\begin{equation}
|{\cal N}, \xi \rangle = \frac{(A^+)^{\cal N}}{\sqrt {{\cal N}!}} |0 \rangle 
= \frac{(B^+)^{\cal N}}{\sqrt {{\cal N}!}} |0 \rangle = |{\cal N}, \alpha \rangle
\, , \quad
\xi_j = \sum^M_{k=1} \frac{e^{i{\tilde k} j}}{\sqrt M} \alpha_k
\label{alxi}
\end{equation}
where the latter definition ensures
$B^+ = \sum^M_{k=1} \alpha_k b^+_k \equiv \sum^M_{j=1} \xi_j a^+_j = A^+$.
Also, the counterpart of formula \ref{Sxi} in the momentum picture is 
easily derived
$$
|{\cal N} ; \alpha \rangle = 
\sum^{({\cal N})}_{\vec p} \,C_{\vec p}({\cal N})\,  
\alpha^{p_1}_1\, ...\alpha^{p_M}_M\, | \vec p \rangle
\, ,
$$
where $C_{\vec p}({\cal N}) := [ {{\cal N}!}/{\prod_k p_k! }]^{1/2}$
while
$|{\vec p} \rangle = [\prod_k p_k !]^{-1/2} \prod_k (b^+_k)^{p_k} | 0 \rangle$
are momentum Fock states. While space-like states $|Z \rangle$ and $|\xi \rangle$
provide information on the local boson population by means of
$\langle Z | a^+_i a_i | Z \rangle= |z_i|^2 $
and $\langle \xi | a^+_i a_i |\xi \rangle= {\cal N} |\xi_i|^2 $, respectively,
momentum-like states $|V \rangle$ and $|\alpha \rangle$ provide information
on the $k$-mode boson population by means of
$\langle V | b^+_k b_k | V \rangle= |v_k|^2 $
and $\langle \alpha | b^+_k b_k |\alpha \rangle= {\cal N} |\alpha_k|^2$,
respectively. The total boson number ${\cal N}$ is unchanged being
$\langle Z | N | Z \rangle = \langle V | N | V \rangle$ and
$\langle \xi | N |\xi \rangle =  \langle \alpha | N |\alpha \rangle$.

As an application of the duality property of $| \xi \rangle$, we
show that states $| S_k \rangle$, describing Schr\"odinger cats, can be defined 
having specific momentum properties. 
To this end an important preliminary condition consists in showing that 
$\cal N$-boson states $|\xi(\ell) \rangle$ with $\ell \in [1,M]$ can be
exploited quite easily to form sets of $M$ orthogonal states.
Recalling that the scalar product of two SU(M) CS is given by
$\langle \eta | \xi \rangle = (\sum_j \eta^*_j \xi_j )^{\cal N}$, one has
$$
\langle \xi (h) | \xi(\ell) \rangle = 
\Bigl ( \sum_j \xi^*_j (h) \xi_j(\ell) \Bigr )^{\cal N} = \delta_{h \ell}
\quad \Leftrightarrow \quad
\sum_j \xi^*_j (h) \xi_j(\ell)= \delta_{h \ell}
$$ 
which
displays how the desired orthogonality directly ensues from the orthogonality
of complex vectors ${\vec \xi} (\ell) = (\xi_2(\ell), \xi_2(\ell), ...)$
with $\ell \in [1,M]$. For fully localized states $|\xi(\ell) \rangle$
characterized by $\xi_j(\ell) \equiv \, \delta_{j\ell}$, 
the orthogonality condition 
$\langle\xi (h)|\xi(\ell)\rangle =\delta_{h \ell}$
is manifest. In the general case, however, the condition 
$\sum_j \xi^*_j (h) \xi_j(\ell)= \delta_{h \ell}$ with
$|\xi_\ell (\ell)| >> |\xi_j(\ell) |$ can be achieved
by exploiting the arbitrariness of the phases of $\xi_j(\ell)$. 
States $|\xi(\ell) \rangle$ describing strong boson localization have been
employed to realize Schr\"odiger-cat states $| S_k \rangle$ 
in a ring of attractive bosons \cite{BPV}. These were proven to well approximate the
low-energy states including the ground state in the regime of strong
interaction. Following the recipe given in \cite{BPV} we define $| S_k \rangle$
as a superposition of equal-weight localized states
$$
| S_k \rangle = \sum^M_{\ell =1} \frac{e^{i{\tilde k} \ell}}{\sqrt M}
|\xi(\ell) \rangle\, ,\quad
|\xi_\ell (\ell)| >> |\xi_j(\ell) | \, , \,\,\, j\ne \ell \, .
$$
As a consequence of the orthogonality of states $|\xi(\ell) \rangle$
states $| S_k \rangle$ appear themselves to be orthogonal namely
$
\langle S_q | S_k \rangle 
= \delta_{qk}
$.
We observe that, if 
$\langle \xi (\ell)| n_i |\xi(\ell)\rangle = {\cal N}|\xi_i (\ell)|^2
\simeq {\cal N} \delta_{i \ell} $ evidences the information about
boson localization at the $\ell$th site, the expectation value
$$
\langle S_k | n_i | S_k \rangle =
\sum^M_{h=1} \sum^M_{\ell=1} 
\frac{e^{i ({\tilde k}\ell-{\tilde q}h) }}{M}\, 
\langle\xi (h)| n_i |\xi(\ell)\rangle
= \frac{ {\cal N} }{M}\, , \quad \forall i\, ,
$$
obtained through the properties
$\langle \eta | a^+_m a_i | \xi \rangle
= {\cal N} \eta^*_m \xi_i \langle \eta |\xi \rangle^{1-1/{\cal N}} $
and $\langle \xi (h)|\xi(\ell)\rangle = \delta_{h \ell} $, confirms the
the expected feature of full delocalization typical
of Schroedinger states.
We note how the possibility of constructing set of orthogonal 
trial states is quite important for applications to boson lattice
systems such as model \ref{BHH}. While trial states can be used
for approximating in a reliable way sets of energy eigenstates,
the possibility to make them mutually orthogonal certainly enriches the 
approximation with an important feature. 

In order to show that states $| S_k \rangle$ have specific momentum properties 
we rewrite $| \xi (\ell) \rangle$ in its dual form
$| \alpha (\ell) \rangle$. Thanks to 
formula \ref{alxi} $| \xi (\ell) \rangle =| \alpha (\ell) \rangle$
with $\alpha_k (\ell) = \sum^M_{j=1} 
{e^{-i{\tilde k} j}} \xi_j(\ell)/{\sqrt M}$, one obtains
$\langle  {\vec p}| \xi (\ell) \rangle =
\langle  {\vec p}| \alpha (\ell) \rangle =
C_{\vec p}({\cal N})\, \alpha^{p_1}_1\, ...\alpha^{p_M}_M$ giving
$$
\langle  {\vec p}| S_q \rangle =
\frac{1}{\sqrt M} \sum_\ell e^{i\ell {\tilde q}}
\langle  {\vec p}| \xi (\ell) \rangle =
\frac{1}{\sqrt M} \sum_\ell e^{i\ell {\tilde q}}
C_{\vec p}({\cal N})\, \alpha^{p_1}_1 (\ell)\, ...\alpha^{p_M}_M (\ell)\, .
$$
In case of strong localization condition $|\xi_\ell (\ell)|\simeq 1 >> |\xi_j(\ell)|$
leads to the approximation $\alpha_k (\ell) \simeq {e^{-i{\tilde k} \ell}}/{\sqrt M}$ and, in particular, to
$$
\langle  {\vec p}| S_q \rangle \simeq
\frac{1}{\sqrt M} \sum_\ell e^{i\ell {\tilde q}}
C_{\vec p}({\cal N})
\frac{e^{-i \ell \sum_k p_k {\tilde k}}}{M^{\sum_k p_k/2}}
=
\frac{\sqrt {{\cal N}!}}{M^{({\cal N}+1)/2}}  
e^{-\frac{1}{2} \sum_k {\rm ln} (p_k !)} 
\sum_\ell 
e^{i \frac{2\pi \ell}{M} \left [ q - \lambda({\vec p}) \right ]}
$$
where $C_{\vec p}({\cal N}) := [{\cal N}! /{\prod_k p_k!} ]^{1/2}$
and
$\lambda({\vec p}) =\sum_k  k \,  p_k = 0, 1, 2, \, ... \,\, {\rm mod} (M)$. 
The latter represents the eigenvalue of the total quasi-momentum operator 
$P = \sigma \sum_k \, k \,  b^+_k b_k$ with $\sigma ={2\pi}/{M}$ 
such that
$P | {\vec p} \rangle = \sigma \lambda({\vec p}) | {\vec p} \rangle$.
It is worth recalling that, in the discrete geometry of ring lattices, the
quasi-momentum properties are described through the displacement operator
$D = \exp[-i\sigma P]$, whose action is displayed
by $D a_\ell D^+= a_{\ell+1}$ and $D b_k D^+ = b_k e^{ik \sigma}$.
Based on equation 
$D |{\vec p} \rangle= e^{-i\sigma \lambda({\vec p})} |{\vec p} \rangle$
Fock states can be organized in $M$ equivalence classes 
labeled by $\lambda({\vec p})$.
Index $q$ in $| S_q \rangle$ therefore characterizes
the quasi-momentum associated with $| S_q \rangle$ since the term
$\sum_\ell \exp \left[ i\ell [ {\tilde q } - \lambda({\vec p})] \right ]$
in $\langle  {\vec p}| S_q \rangle$ vanishes whenever $| {\vec p} \rangle$ has
a momentum $\lambda({\vec p}) \ne q \,\, {\rm mod} (M)$. States $| S_q \rangle$,
in the presence of strong localization, supply a set of $M$ orthogonal states
whose label $q$ bears information on the class with quasi-momentum $\lambda$
maximally involved in the realization of $| S_q \rangle$, within the
$\cal N$-particle Hilbert space. As for Glauber-like states, we note that
$\langle X |Z \rangle =$ $ \prod_j \langle x_j |z_j\rangle= 
\prod_j \exp [{\bar x}_j z_j -(|z_j|^2 +|x_j|^2)/2] $
so that $|Z \rangle$ and $|X \rangle$ cannot be orthogonal. At most, $M$ quasi-orthogonal states can be obtained by considering sets $\{ x_j(\ell) \}$
such that $|x_j(\ell)|^2 \simeq {\cal N} \delta_{j\ell}$ for which
$|\langle X(h) |X(\ell) \rangle| \simeq e^{-{\cal N}}$. 
Representation of Schr\"odinger-cat states in terms of quasi-orthogonal
states $|Z \rangle$ can be developed under these conditions.

\section{Conclusions}
\label{conc}

In this paper we have compared variational schemes based 
on trial states $|F \rangle$, $|Z \rangle$ and $|\xi \rangle$,
widely used in applications to many-mode boson systems. To illustrate
their distinctive features we have applied such schemes to the BH model
which has became, in the recent years, the paradigm of real 
interacting-boson lattice systems.
Such a comparison has been aimed at evidencing 
the specific characters of each scheme to favour their applications
in the study of the properties of many-mode boson systems.

In section \ref{sez2}, we have applied the
$|F\rangle$-based scheme to the BH model showing, within the
corresponding dynamical scenario, that collective variables
$\alpha_i$, $\alpha^*_i$ form a (classical) Weyl-Heisenberg sub-algebra
in the Poisson algebra of variables
${f}^{j}_n$, ${\bar f}^{\ell}_m$. This crucial property allows 
reduction of the $|F\rangle$-based picture, exhibiting
a more pronounced quantum character, to the $|Z\rangle$-based picture
based on Glauber's CS. The semiclassical character
of the latter appears to be an effective, dynamically-consistent procedure
incorporating the Bogoliubov approximation.

In section \ref{sez3}, we have shown that Glauber-like trial
state $|Z \rangle$
is a superposition of SU(M) CS $|{\cal N}, \xi \rangle$ that involves 
all the ${\cal N}$-particle sectors of the Hilbert space. We have exploited 
this property to explain why the dynamical equations relevant to the BH model
obtained in the $|{\cal N}, \xi \rangle$-based scheme coincide with the
equations derived in $|Z \rangle$-based scheme. The meaning of microscopic
CS parameters of such schemes has been illustrated and related to the fact
that states $|{\cal N}, \xi \rangle$ are boson-number preserving.
Also, in section \ref{sez3}, we have discussed explicitly the procedure that
enables one to recast state \ref{suMCS} into the standard form
$|\xi \rangle = g | \Omega \rangle$ of CS theory
involving the extremal vector $| \Omega \rangle$.

Section \ref{sez4} has been devoted to discuss the duality property
of space-like states $|Z \rangle$ and $|\xi \rangle$ which allows
to rewrite them as momentum-like states involving modes $b^+_k$.
This property has been used for constructing Schr\"odinger-cat states 
with specific momentum features for bosons in ring lattices.
In general, the use of states $|\xi \rangle$ and the possibility
to construct set of orthogonal states outlined in 
section \ref{sez4} should allow a better
characterization of low-energy regimes in systems
of bosons confined in ring lattices whose standard description
is given in terms of Hamiltonian \ref{BHH}. Particularly,
the duality property of states $|Z \rangle$ and $|\xi \rangle$
finds a natural application in the study of
supercurrents and vortex states occurring in such 
systems \cite{DH}, \cite{LAK}.

As observed in section \ref{sez0}, while the $|Z \rangle$-based
scheme has been extensively used in applications,
the interest for the $|\xi \rangle$ and $|F \rangle$-based schemes
is more recent. The more pronounced
quantum character of the $|F \rangle$-based scheme is expected to supply,
in the applications to BH-like models, a better description of the
critical behaviors \cite{SAC}-\cite{lasp} inherent in quantum phase transitions.
For the same reason it should supply an effective tool to study the
complex dynamics \cite{JVC}, \cite{DST} of bosons in lattice systems.
Bosons distributed in small arrays (and thus involving small number $M$
of space modes) are specially interesting since they can switch from fully
quantum to intermediate semiclassical behaviors by adjusting model parameters
\cite{LAK}-\cite{BPV}. The correspondingly small number of components
$|F_i \rangle$ in $|F \rangle$ makes it feasible for performing numerical
simulations of equations \ref{Geq}. These aspects will be investigated
in a separate paper \cite{BMPV}.

%
\begin{appendix}

\section{Application of formula \ref{suMCS} to the case M=2}
\label{M2}
It is quite easy to show that formula \ref{suMCS} with
$M=2$ reproduces the standard group-theoretic definition of SU(2) coherent state. In this case $A^+ = \xi_1 a^+_1+ \xi_2 a^+_2$. Then
$$
|{\cal N}, \xi \rangle =
({{\cal N}!})^{-1/2}\, (A^+)^{\cal N} |0, 0 \rangle=
\frac{1}{\sqrt {{\cal N} !}} \sum^{{\cal N}}_{s=0}
\frac{{\cal N}! \, \xi_2^s \xi_1^{{\cal N} -s} }{ s! ({\cal N} -s)! } \,
(a^+_2)^s (a^+_1)^{{\cal N} -s} |0, 0 \rangle
$$
$$
=\sum^{{\cal N}}_{s=0}
C_s({\cal N}) 
%
%
\,
\xi_2^s \xi_1^{{\cal N} -s} |{\cal N} -s, s \rangle
=
e^{i {\cal N} \phi_1}
\sum^{{\cal N}}_{s=0}
\frac{C_s({\cal N}) z^s }{(1+|z|^2)^{{\cal N}/2}} \, |J; -J+s \rangle
= e^{i {\cal N} \phi_1} |J; z \rangle
$$
where $C_s({\cal N}) \equiv 
\sqrt{{\cal N}!}/ \sqrt{ s! ({\cal N} -s)! }$, the
definition $z= \xi_2/\xi_1$ has been used, and $\phi_\ell$ is the phase
of $\xi_\ell $. Moreover,
$|J; -J+s \rangle \equiv |{\cal N} -s, s \rangle$, where $J= {\cal N}/2$, 
can be seen as the
$m$th vector (with $m=-J+s$) in the standard basis
$\{ |J; m \rangle: J_3 |J; m \rangle= \, m |J; m \rangle \}$ 
of algebra su(2) within the Schwinger boson picture 
of spin operators
$J_3 = (a^+_2 a_2 - a^+_1 a_1)/2$, $J_+ = a^+_2 a_1= (J_-)^+$.
State
$g |\Omega \rangle = e^{\zeta J_+ - \zeta^* J_-} |J; -J \rangle
= (1+|z|^2)^{-J} e^{ z J_+} |J; -J \rangle $, obtained through
the standard decomposition \cite{ZFG} 
$g = e^{\zeta J_+ - \zeta^* J_-}=
e^{z J_+}$ $ e^{J_3 {\rm ln} (1+|z|^2)} e^{-z^* J_-}$,
coincides with state $|J; z \rangle$ just defined,
where $z$ and $\zeta$ have the same phase 
and $|\zeta|= {\rm tg} |z|$.

\section{Derivation of dynamical equations for $z_j$}
\label{glauber}

Based on dynamical equations \ref{Geq} governing the evolution of 
variables $f^{i}_m$ one has
$$
i \frac{d \alpha_i}{dt} 
 = {\sum}_m {\sqrt {m +1}} \, 
\left [ 
i \frac{d {\bar f}^{i}_m }{dt} f^{i}_{m+1}
+
i \frac{d f^{i}_{m+1} }{dt}  {\bar f}^{i}_m
\right ]
$$
$$
=
{\sum}_m {\sqrt {m +1}} \, 
\Bigl [ 
f^{i}_{m+1}
\left (
-\frac{_U}{^2} (m^2-m) {\bar f}^{i}_m
+ {\sqrt {m+1}} {\bar f}^{i}_{m+1} \, \Phi_i 
+ {\sqrt {m}} {\bar f}^{i}_{m-1} \, \Phi^*_i
\right )
+
$$
$$
+
{\bar f}^{i}_m
\left (
\frac{_U}{^2} (m^2+m) f^{i}_{m+1}
-
{\sqrt {m+2}} f^{i}_{m+2} \, \Phi^*_i - {\sqrt {m+1}} f^{i}_{m} \, \Phi_i
\right ) \Bigr ]
$$
$$
={\sum}_m \, U m {\sqrt {m +1}} \, {\bar f}^{i}_m f^{i}_{m+1}
+
\Phi_i 
\left [ 
{\sum}_m
\left (
(m+1) {\bar f}^{i}_{m+1} f^{i}_{m+1} - m\, {\bar f}^{i}_m f^{i}_{m} 
\right )
-
{\sum}_m {\bar f}^{i}_m f^{i}_{m}
\right ]
+
$$
$$
+
\Phi^*_i 
\left (
{\sum}_m
{\sqrt {m +1}} {\sqrt {m}} {\bar f}^{i}_{m-1} f^{i}_{m+1}
- 
{\sum}_m
{\sqrt {m +1}} {\sqrt {m+2}} {\bar f}^{i}_m f^{i}_{m+2} 
\right ) \, ,
$$
where the index-$m$ range is $[0,\infty]$.
Thus ${\dot \alpha}_i$ is formed by three terms.
Substitution $m \to m+1$ in the first summation of the third term
(one should notice that ${\bar f}^{i}_{-1} =0$) shows that the latter 
vanishes, while, in the second term,
$
\Sigma_{m}
\left [ (m+1) {\bar f}^{i}_{m+1} f^{i}_{m+1} 
- m\, {\bar f}^{i}_m f^{i}_{m} 
\right ]
= 0
$ is easily proven.
Further simplification is achieved if the
the on-site normalization condition
$\langle F_i|F_i \rangle = 
\sum_{m} {\bar f}^{i}_m f^{i}_{m}=1$ is imposed.
Notice that choice \ref{assum} authomatically ensures
such a condition since $|F_i \rangle = |z_i \rangle$ and
Glauber CS are such that $\langle z_i|z_i \rangle=1$.
Under this circumstance the second term reduces to $\Phi_i$.

\section{Conservation of ${\cal N}$ and other constants of motion}
\label{due}

After recalling that 
${\cal N}=\langle \Psi | N |\Psi \rangle
=
\langle \Psi | \sum_j n_j | \Psi \rangle
=
{\sum}_j {\sum}_n \, n |f^{j}_n|^2$
where $\langle \Psi | n_j |\Psi \rangle ={\sum}_n n |f^{j}_n|^2$
and $n \in [ 0,\infty ]$, $j \in [1,M]$,
let us consider the Poisson bracket of ${\cal N}$ and $\cal H$
$$
\{ {\cal N}, {\cal H} \}
=
{\sum}_j {\sum}_n\,  n\, \{ |f^{j}_{n}|^2, {\cal H} \}
=
{\sum}_j {\sum}_n\,  n 
\left [ {\bar f}^{j}_n \{ f^{j}_{n}, {\cal H} \} 
+ f^{i}_n \{ {\bar f}^{i}_n, {\cal H} \} \right ]
=
$$
$$
=
-i {\sum}_j {\sum}_n\,  n 
\Bigl [ {\bar f}^{j}_n 
\left ( \frac{_U}{^2} (n^2-n) f^{j}_n
-
{\sqrt {n+1}} f^{j}_{n+1} \, \Phi^*_j - {\sqrt n} f^{j}_{n-1} \, \Phi_j
\right )
$$
$$
+ f^{j}_n 
\left (
-\frac{_U}{^2} (n^2-n) {\bar f}^{j}_n
+
{\sqrt {n+1}} {\bar f}^{j}_{n+1} \, \Phi_j + {\sqrt n} {\bar  f}^{j}_{n-1} 
\, \Phi^*_j
\right ) \Bigr ]
$$
$$
\quad\,\, = 
-i\, {\sum}_j \Phi^*_j {\sum}_n \, 
\left ( {\sqrt {n+1}}  f^{j}_{n+1} {\bar f}^{j}_{n}
\right )  
-i\, {\sum}_j \Phi_j {\sum}_n \, 
\left ( - {\sqrt {n+1}} {\bar f}^{j}_{n+1} f^{j}_{n}
\right )
$$
$$
\,\, = 
-i\, {\sum}_j \Phi^*_j \alpha_j -{\sum}_j \Phi_j \alpha^*_j
=
-i \, {\sum}_j  \alpha_j  {\sum}_{\ell\in j} 
\left ( T_{\ell j} \alpha^*_\ell- T_{\ell j} \alpha_\ell \right ) =0 \, .
$$
Then $d{\cal N}/dt= \{ {\cal N} , {\cal H} \} = 0$.
A similar calculation allows one to evidence that
other $M$ constants of motion are involved in the
dynamics of $f^j_n$. These are
$I_j= {\sum}_n |f^j_n|^2$ for which $ \{ {\cal I}_j , {\cal H} \} = 0$.
Quantities $I_j$ are in involution with $\cal N$, $\alpha_i$ and $I_i$ 
namely $\{ I_j , {\cal N} \} = 0$ and $\{ I_j , \alpha_i \} =0$ $\forall j$.
One easily check as well that $\{ I_j , I_i \} =0$ for each $i$ and $j$.

\section{Formulas relevant to the SU(M)-CS picture}
\label{tre}

In order to obtain
${\cal L} (\xi) = 
i\hbar \langle \xi |\partial_t | \xi \rangle -\langle \xi | H| \xi \rangle$
one needs to calculate $\langle \xi |\partial_t | \xi \rangle$ and
$\langle \xi| (n_i-1) n_i |\xi \rangle$ in 
${\cal H}(\xi) =\langle \xi | H| \xi \rangle$.
Concerning $\langle \xi| (n_i-1) n_i |\xi \rangle$, 
(recall that $|\xi \rangle \equiv |S ; \xi \rangle$) one should observe
that $[a_i, (A^+)^s] = s \xi_i\, (A^+)^{s-1}$, 
and that
$$
a_i |\xi \rangle
=
a_i \rho_S (A^+)^S |0 \rangle
=\rho_S [ (A^+)^S a_i + S \xi_i (A^+)^{S-1} ] |0 \rangle
={\sqrt S} \xi_i|\xi' \rangle
$$
where $\rho_S = {1}/{\sqrt {S!}}$,
and $|\xi' \rangle= \rho_{S-1} (A^+)^{S-1} |0 \rangle$
is a $(S-1)$-boson coherent state. Iterating this calculation gives
$a^2_i |\xi \rangle = \xi^2_i {\sqrt {S (S-1)} } \, | \xi ''\rangle$
with $| \xi ''\rangle = \rho_{S-2} (A^+)^{S-2} |0 \rangle $.
Therefore $\langle \xi| n_i |\xi \rangle =\, ...= N\, |\xi_i|^2$ 
whereas
$$
\langle \xi| (n_i-1) n_i |\xi \rangle
=\langle \xi| (a^+_i)^2 a^2_i |\xi \rangle
=
S(S-1)\, |\xi_i|^4  \, \langle \xi'' | \xi'' \rangle =S(S-1)\, |\xi_i|^4
\, .
$$
Similarly, one finds
$\langle \xi| a^+_m a_i |\xi \rangle
= \langle 0 | \rho_S A^S  
a^+_m \, a_i \rho_S {(A^+)^S } |0 \rangle
=
S \xi_m^* \xi_i \, \langle \xi' | \xi' \rangle
=S \xi_m^* \xi_i$.
For two generic states $|\eta \rangle$, $|\xi \rangle$
the latter becomes $\langle \eta | a^+_m a_i |\xi \rangle = 
S \eta_m^* \xi_i \, 
( \sum_i \eta_i^* \xi_i )^{S-1}$ where the inner product 
$\langle \eta | \xi \rangle= ( \sum_i \eta_i^* \xi_i )^{S}$ of
two $S$-boson states $|\eta \rangle$, $|\xi \rangle$ 
has been used.
In the effective Lagrangian ${\cal L}$ term
$\langle \xi | \partial_t |\xi \rangle$ can be recast as
$$
\langle \xi | \partial_t |\xi \rangle
=
\langle \xi | {\sum}_j {\dot \xi}_j { \partial_{\xi_j}} |\xi \rangle
=
\rho^2_{S-1} \langle 0 | A^{S-1}
{\sum}_j {\dot \xi}_j A a^+_j (A^+)^{S-1} |0 \rangle
={\langle \xi' |} {\sum}_j {\dot \xi}_j A a^+_j  |\xi' \rangle
$$
$$
=
\langle \xi' | {\sum}_j {\dot \xi}_j \left [ a^+_j A +\xi^*_j \right ]
|\xi' \rangle
= 
{\sum}_j {\dot \xi}_j \xi^*_j +
{\sum}_j {\sum}_m \xi^*_m {\dot \xi}_j 
\, \langle \xi' | a^+_j a_m |\xi' \rangle
$$
$$
= {\sum}_j {\dot \xi}_j \xi^*_j 
+ {\sum}_j {\sum}_m \xi^*_m {\dot \xi}_j 
\, (S-1) \xi_m \xi^*_j
=
{\sum}_j {\dot \xi}_j \xi^*_j +(S-1) {\sum}_j  {\dot \xi}_j \xi^*_j 
= S {\sum}_j  {\dot \xi}_j \xi^*_j
\, .
$$
Concluding,
the four/two-boson expectation values just obtained
provide the effective Hamiltonian
$\langle \xi | H |\xi \rangle= 
\frac{U}{2} S(S-1) \sum_j  |\xi_j|^4 - 
S \, \sum_{<j,\ell>} T_{j\ell} \xi_j^* \xi_\ell $
occuring in ${\cal L}(\xi)$.

\section{Derivation of operator $A^+$}
\label{quattro}

After setting $\sum^M_{k=2} \theta^2_k : = \theta^2$, the action of 
$e^{iD}$ on $a^+_1$ is given by the standard formula
$$
e^{iD} a^+_1 e^{-iD} = \sum^{\infty}_{s=0} \frac{i^s }{s!} [D, a^+_1]_s
$$
where 
$[D, a^+_1]_s = [D, [D, a^+_1]_{s-1} ]$, $[D, a^+_1]_1= [D, a^+_1]$, and 
$[D, a^+_1]_0 = {\bf 1}$. Observing that $[D, a^+_1]_{2r} = \theta^{2r}  a^+_1$
and
$[D, a^+_1]_{2r+1} = \theta^{2r} \, Q$
with $Q = \sum^M_{k=2} \theta_k  a^+_k$, one obtains
%
%
$$
e^{iD} a^+_1 e^{-iD} = 
\sum^{\infty}_{r=0} \frac{(-)^r }{(2r)!} \theta^{2s}  a^+_1
+
\sum^{\infty}_{r=0} \frac{i (-)^r }{(2r+1)!} \theta^{2s} Q
=
a^+_1 \cos \theta +i Q \frac{\sin \theta}{\theta}
=
\sum^M_{j=1} y_j  a^+_j
$$
where $y_1 = \cos \theta$ and 
$y_k= i\theta_k {\sin \theta}/{\theta}$ if $ k \ne 1$.

\section{Two-boson operators of algebra su(M)}
\label{uno}

The fact that $T(\zeta) \in$ SU(M) is easily demonstrated by recalling 
that, within a Schwinger-like picture, algebra su(M) can be realized in terms
of two-boson operators
$a^+_j a_k$, $ a^+_k a_j$ with $1 \le j \le M-1$ and $j+1 \le k \le M$
that play the role of lowering and raising operators, respectively.
This set is completed by the generators
of the Cartan-subalgebra $\{ h_k, k= 2,... M: [h_k, h_\ell ]=0 \}$
where each of the
$M-1$ operators $h_k$ can be written as an appropriate linear combinations
of number operators $n_i = a^+_i a_i$, $i= 1,2...\, , M$.
We notice that, consistent with the presence of the group-invariant
operator $N =\sum^{M}_{i= 1} n_i$, only $M-1$ operators $h_k$ can be
realized with $M$ operators $n_i$. A generic element of 
${\cal G}$ = su(M) $= \{ a^+_j a_m \, (m \ne j) : \,
m, j \in [1,M]; \, h_k\, : k \in [2,M] \}$
is thus given by
$$
\sum^{M-1}_{j=1} \sum^{M}_{k= j+1} ( z_{kj} a^+_j a_k + z^*_{kj} a_j a^+_k )
 +\sum^{M}_{k= 2} \, \alpha_k \, h_k
$$
where $z_{kj} = x_{kj} + i y_{kj}$ and
$x_{kj}$, $y_{kj}$, $\alpha_k \, \in {\bf R}$. Since
elements $g \in G$ of a Lie group $G$ are generated by the
Lie algebra element $a \in {\cal G} = Lie (G)$ 
through the exponential map $g = \exp ( i a)$ then the 
latter fromula shows that $T(\zeta) \in $ SU(M).

\end{appendix}

\section*{References}

\begin{thebibliography}{99}
\bibitem{com1} 
Even if $|Z \rangle$ and $|F \rangle$
do not conserve $N$, this property is recovered within the mean-field picture 
where $\langle \Phi | N | \Phi \rangle $ with $\Phi = Z, F$
is a constant of motion of the effective Hamiltonian 
${\cal H} = \langle \Phi| H | \Phi \rangle$.
\end{thebibliography}

\end{document}